\documentclass[sigconf,nonacm]{acmart}
\setcopyright{none}
\usepackage{booktabs}
\usepackage{xurl}
\usepackage{tikz}
\usetikzlibrary{arrows.meta}
\newcommand{\systemname}{CATP}

\title{CATP: Design and Evaluation of Local Agent Authorization and Audit Evidence}

\author{Linfeng Zhou}
\affiliation{\institution{Stealth}\country{}}

\begin{document}

\begin{abstract}
A signed authorization record authenticates what a signer asserted, but does
not by itself establish that the assertion agrees with the policy and action
used for a runtime decision. \systemname{} specifies the bindings needed to
carry a local pre-execution decision into offline-verifiable evidence. Its hook
commits to the enforcement-time policy and adapter-normalized action and durably
records the decision before returning permission. A later receipt binds the
record to a self-contained audit export.
An implementation for Claude Code and Codex supports an empirical examination
of these bindings. Controlled, validly signed counterexamples expose five
missing consistency checks; adding them changes acceptance on the same inputs,
with all 13 expected outcomes satisfied. A separate probe finds that distinct
argument vectors can become the same normalized action before hashing, limiting
what even a consistent receipt can establish. Eight scripted runtime cases
verify hook delivery and allow/deny behavior; failure tests cover tampering,
substitution, replay, concurrency and persistence errors. Median paired hook
overhead is approximately 85\,ms, and offline verification takes approximately
174\,ms end to end on the testbed. The evidence remains conditional on a trusted
runtime, monitor, host and signing key. It authenticates a decision about the
normalized action, without proving execution truth, complete disclosure or
autonomous task utility. The optional Groth16 study covers a narrower circuit
predicate; the evaluation does not establish superiority over other systems.
\end{abstract}

\maketitle
\raggedbottom

\section{Introduction}

Consider an agent that requests a repository file edit under a local access
policy. After the request, an auditor receives a signed record saying that the
edit was allowed. The signature authenticates the record, but leaves important
questions unanswered: does it bind the complete edit request or only a summary?
Does it identify the policy used at the decision point or one loaded later?
Was evidence durably recorded before the runtime received permission? These
are different requirements from detecting edits to a log.

\systemname{}\footnote{Source code: \url{https://github.com/lfzkoala/catp}.}
records the policy and action used for a pre-execution decision
so that an auditor can later check the evidence offline. The enforcement path
snapshots the policy and canonical action, evaluates the request, and durably
records their commitments with the decision before returning permission. A
receipt issued later binds the selected pre-execution entry to a self-contained
audit export. The verifier checks the signature, export, chain, selected entry
and complete adapter-normalized action. It can also check a supplied policy
against the recorded policy commitment. The runtime must deliver the hook and
obey a denial; the monitor, operating system and signing key remain trusted.
These assumptions are necessary because the receipt cannot prove execution on
an untrusted host.

Reference monitors, commitments, hash chains and signatures are established
techniques. SAGA provides cryptographic authorization for agent
interactions~\cite{syros2026saga}, while Signet and the Permit draft describe
authorization evidence for agent actions~\cite{signet2026,munoz2026permit}.
CATP addresses the implementation constraints at a local runtime hook: which
policy and action to commit, when to persist them, and how to check their
agreement with a later receipt. Its contract binds the enforcement-time policy,
adapter-normalized action, decision phase, audit position and signer identity.
Permission depends on durable recording, and receipt verification requires the
corresponding audit export.

The implementation connects Claude Code and Codex to a common enforcement and
receipt pipeline. It provides a way to examine where the chain from a runtime
request to signed evidence can lose meaning. One failure occurs at verification:
a signed decision can disagree with the audit entry it names if the verifier
does not compare the copied fields. Another occurs before signing: an adapter
can map distinct argument vectors to the same normalized action. Checking every
receipt field resolves the first problem but cannot recover information already
lost during adaptation. Different tool names and input fields also limit policy
portability between runtimes.

The evaluation uses controlled receipt counterexamples, adapter probes,
durability tests and paired timing measurements. Five validly signed
contradictions expose omitted field comparisons in the implementation; checking
those fields rejects the same inputs. Artifact identities and the correction
are recorded in Appendix~\ref{sec:artifact-provenance}. All eight scripted
shell/file allow/deny cases pass on the real runtimes. These cases test hook
behavior, not autonomous task completion. Median paired enforcement overhead is
approximately 85\,ms on our testbed, and receipt verification takes approximately
174\,ms end to end. The measurements cover full paths; they neither isolate
component costs nor establish superiority over other systems. Separate studies
examine the optional Groth16 authorization circuit and external anchoring. The
circuit checks a narrower predicate, while an independently retained checkpoint
can expose deletion of covered log entries. Neither establishes execution truth
or complete disclosure.

\section{Background and motivation}

Agent runtimes expose tools such as shell execution and file modification, but
their event payloads and blocking semantics differ. A common authorization
protocol needs an action representation that each adapter can produce. That
shared representation does not make the adapters' mediation guarantees identical.

An allow or deny response is ephemeral. To check it later, an external verifier
needs evidence identifying the policy, action, decision phase, audit record and
signer. The verifier must obtain the trusted public key independently; an
embedded key alone cannot establish who is authorized to sign. Given a runtime
event and a local policy, CATP must produce evidence that supports this check
within explicit assumptions about the runtime and host.

Plain mutable logs lack independently checkable authentication. Signing an
arbitrary description authenticates its bytes, but cannot recover omitted action
fields, identify a policy loaded earlier or establish persistence order. A
signed audit system can meet CATP's contract by recording the same policy
snapshot and full normalized action, enforcing the persistence barrier and
validating the same bindings. The implementation must enforce those constraints
in addition to checking the signature.

\section{Threat model and security goals}

\subsection{Adversary}

We consider an agent that can choose tool names and inputs, including malformed
or adversarial payloads, and can attempt to exploit policy matching, replay old
evidence, substitute a policy, or alter individual audit and receipt fields. We
also consider concurrent tool invocations that race while appending audit
entries.

\subsection{Trust assumptions}

The runtime must deliver the documented pre-execution event to the adapter and
honor its blocking result. The verifier must possess an independently trusted
receipt public key. The local operating system, CATP executable, and signing-key
custody remain part of the trusted computing base unless protected by an
external mechanism.

\subsection{Security goals}

\begin{itemize}
  \item \textbf{Supported hook mediation:} every
        pre-execution event that an adapter normalizes into a canonical action is
        evaluated before the runtime executes it. This guarantee is bounded to
        the tool-action paths the adapters actually surface and that the
        evaluation enumerates and tests (Section~\ref{sec:runtime-coverage}); it does not extend to events
        an adapter fails to normalize into the shape the policy engine matches.
  \item \textbf{Fail-closed enforcement:} malformed events, invalid policies,
        and audit persistence failures do not silently authorize an action once
        policy enforcement is active. This does not cover a valid payload whose fields
        do not match a policy predicate; an unmatched rule can default to allow.
  \item \textbf{Binding of normalized evidence:} changing a bound policy, action, phase,
        decision, audit entry, or signer without a valid new signature causes
        verification to fail. Consistency of validly signed copied fields is a
        separate requirement tested with validly signed counterexamples
        in Section~\ref{sec:mechanism}.
  \item \textbf{Audit consistency:} accepted records form one ordered commitment
        chain, including under concurrent append attempts.
  \item \textbf{Truncation visibility (conditional):} with optional asynchronous
        anchoring enabled, deletion of an anchored audit suffix is detectable by
        any holder of the published anchor. The uncovered suffix begins at the
        last successfully published and independently retained checkpoint; under
        the local-only default, suffix truncation is detectable only from an
        independently retained receipt or anchor.
\end{itemize}

\subsection{Non-goals}

\systemname{} does not prove that a policy is semantically correct, that an
allowed action is harmless, that a runtime executed the reported action, or
that a fully compromised host preserved evidence. Under the local-only default,
suffix truncation is not detectable from the chain alone; enabling optional
anchoring exposes deletion of checkpointed records but adds the external witness
to the trust assumptions. A configured interval does not guarantee publication
liveness or a wall-clock detection bound.

\section{Protocol design}

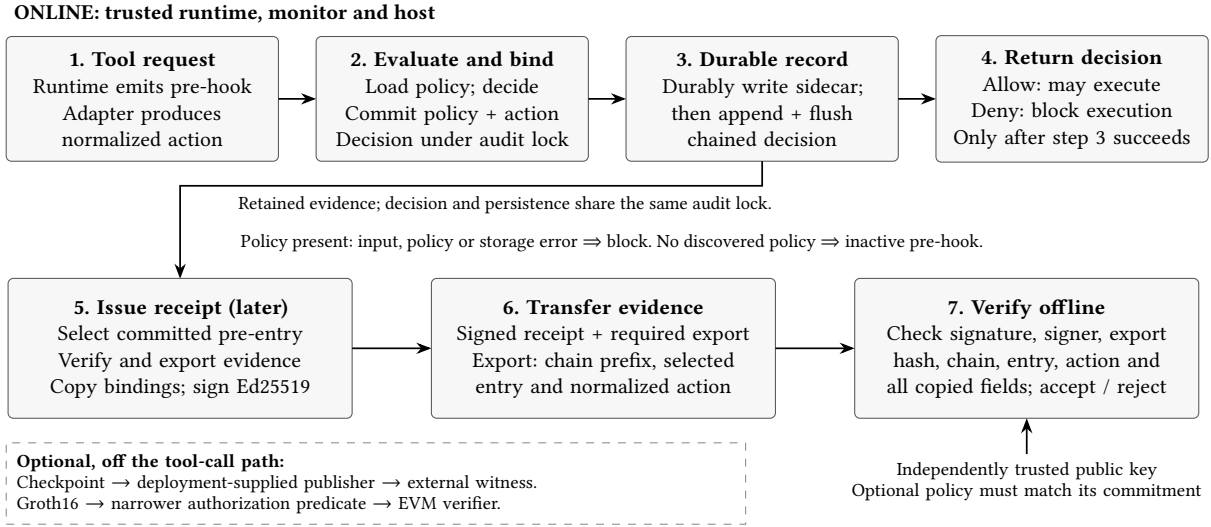
\begin{figure*}[t]
\centering
\begin{tikzpicture}[
  font=\small,
  protocolbox/.style={draw=black!65, rounded corners=2pt, fill=black!3,
    align=center, inner sep=5pt, minimum height=1.65cm},
  protocolarrow/.style={-{Stealth[length=2mm]}, line width=.65pt},
  protocolnote/.style={align=left, font=\footnotesize},
  every node/.style={outer sep=0pt}
]
\node[anchor=west,font=\small\bfseries] at (0,1.13)
  {ONLINE: trusted runtime, monitor and host};
\node[protocolbox,text width=3.25cm] (request) at (1.8,0)
  {\textbf{1. Tool request}\\Runtime emits pre-hook\\Adapter produces\\normalized action};
\node[protocolbox,text width=3.25cm] (evaluate) at (5.9,0)
  {\textbf{2. Evaluate and bind}\\Load policy; decide\\Commit policy + action\\Decision under audit lock};
\node[protocolbox,text width=3.25cm] (persist) at (10,0)
  {\textbf{3. Durable record}\\Durably write sidecar;\\then append + flush\\chained decision};
\node[protocolbox,text width=3.25cm] (release) at (14.1,0)
  {\textbf{4. Return decision}\\Allow: may execute\\Deny: block execution\\Only after step 3 succeeds};
\draw[protocolarrow] (request) -- (evaluate);
\draw[protocolarrow] (evaluate) -- (persist);
\draw[protocolarrow] (persist) -- (release);
\draw[protocolarrow] (persist.south) -- (10,-1.15) -- (2.3,-1.15) -- (2.3,-2.38);
\node[protocolnote,anchor=west,fill=white,inner sep=2pt] at (3.0,-1.40)
  {Retained evidence; decision and persistence share the same audit lock.};
\node[protocolnote,anchor=west] at (3.0,-1.90)
  {Policy present: input, policy or storage error $\Rightarrow$ block. No discovered policy $\Rightarrow$ inactive pre-hook.};
\node[protocolbox,text width=4.2cm,minimum height=1.85cm] (issue) at (2.3,-3.30)
  {\textbf{5. Issue receipt (later)}\\Select committed pre-entry\\Verify and export evidence\\Copy bindings; sign Ed25519};
\node[protocolbox,text width=4.2cm,minimum height=1.85cm] (bundle) at (7.9,-3.30)
  {\textbf{6. Transfer evidence}\\Signed receipt + required export\\Export: chain prefix, selected\\entry and normalized action};
\node[protocolbox,text width=4.2cm,minimum height=1.85cm] (verify) at (13.5,-3.30)
  {\textbf{7. Verify offline}\\Check signature, signer, export\\hash, chain, entry, action and\\all copied fields; accept / reject};
\draw[protocolarrow] (issue) -- (bundle);
\draw[protocolarrow] (bundle) -- (verify);
\node[protocolnote,draw=black!45,dashed,inner sep=4pt,text width=9.5cm,anchor=north west]
  at (0,-4.55) {\textbf{Optional, off the tool-call path:}\\
  Checkpoint $\to$ deployment-supplied publisher $\to$ external witness.\\
  Groth16 $\to$ narrower authorization predicate $\to$ EVM verifier.};
\node[protocolnote,align=center,text width=4.7cm,anchor=north] (key) at (13.5,-4.70)
  {Independently trusted public key\\Optional policy must match its commitment};
\draw[protocolarrow] (key.north) -- (verify.south);
\end{tikzpicture}
\caption{Protocol overview. Durable recording precedes permission; receipt issuance
is later. Offline verification authenticates supplied evidence under trusted
enforcement and signing components, not physical execution, host honesty or
complete disclosure. Optional backends retain their narrower scope.}
\label{fig:protocol-overview}
\end{figure*}

Figure~\ref{fig:protocol-overview} separates the synchronous authorization path
from subsequent evidence issuance and independent verification.

\subsection{Runtime-neutral actions}

Adapters map runtime payloads into a common action containing the runtime,
phase, tool name, structured input, and optional session metadata. Runtime-native
tool names are retained so the policy and audit record describe the surface the
runtime actually exposed. Canonical action binding starts after adaptation.
It cannot recover information discarded by an adapter: the evaluated Codex
adapter joins argv elements with spaces, losing argument boundaries
(Section~\ref{sec:adapter-boundary}). Thus ``complete canonical action'' means
the complete normalized object, not an injective encoding of raw requests.

\subsection{Policy evaluation}

Rules are evaluated in order and the first match determines the result. Policy
discovery activates the enforcement boundary: once a policy is present, input,
policy, or persistence failures block rather than bypass the decision.

\subsection{Commitment-chained audit}

Every commitment is SHA-256 over a fixed ASCII domain string concatenated with
canonical (key-sorted) JSON of the committed object. Domain strings end in a
newline byte. Domain separation
distinguishes commitment roles, subject to SHA-256 collision resistance.
The action domain is \texttt{catp:action:v1}; the policy domain is
\texttt{catp:policy:v1}. Canonical action evidence includes runtime, phase,
tool name, adapter-normalized structured input and optional session id. It excludes the
runtime-native raw payload, which can carry irrelevant data or secrets.
Each audit
entry is a version-4 record (\texttt{commitment\_version}=4) whose commitment,
under domain \texttt{catp:audit-entry:v4}, covers the phase, tool, decision,
timestamp, matched rule, reason, input summary, the enforcement-time policy and
action commitments, an optional authorization block, and the previous entry's
commitment, so each entry commits to its predecessor and to the stored
authorization fields. Appends are serialized per
agent and day so concurrent hook processes cannot create sibling chain heads.

\subsection{Signed authorization receipts}

A version-2 receipt (\texttt{catp\_\allowbreak authorization\_\allowbreak receipt\_v2}) is an Ed25519
signature, under domain \texttt{catp:receipt-signature:v2}, over a body binding
the selected pre-execution audit entry and its commitment, the
\texttt{catp:audit-export:v2} hash of a self-contained export, the policy and
action commitments, agent and tool identifiers, the decision, timestamp, and the
signer key id (SHA-256 of the public key's DER SubjectPublicKeyInfo). The policy
and action commitments are \emph{copied} from the enforcement-time v4 entry, never
re-derived from the current policy file, so a receipt attests to what was decided
at enforcement time. The self-contained export contains the daily chain prefix
through the selected entry and the complete canonical action. It establishes
the entry's place in the supplied chain and binds it to that normalized action.
Later entries and undisclosed logs fall outside this evidence; the signature
covers only the selected receipt body. Version-2 verification therefore requires
the audit export as well as a valid signature. The verifier checks the
export hash, chain, selected entry, action commitment, and full canonical action
before returning enforcement-time-bound assurance. The contract additionally
requires every copied receipt field to equal the selected entry.
Section~\ref{sec:mechanism} tests the effect of omitting these comparisons.
A policy file is optional
additional evidence for checking the policy commitment.

\paragraph{Enforcement-to-evidence invariants.}
The policy and canonical action used for a decision must be the ones committed
in its pre-execution record. Successful permission requires durable action
sidecar and audit persistence; persistence failure must not release the action.
Receipt issuance is subsequent to this critical path and preserves these
commitments rather than reconstructing them from mutable policy files. These
constraints link a decision to evidence under the trusted implementation; the
receipt cannot independently prove the host's execution order or honesty.

\subsection{Optional proof backend}

The authorization statement is versioned independently of its verifier. The
default path uses signed receipts. The optional Groth16 backend over BN254 implements
a narrower circuit-defined authorization predicate for EVM verification, not
the full TOML policy language or the complete receipt/export verification
contract. Proof generation is not required for local enforcement.

\subsection{Optional asynchronous anchoring}

A local chain cannot prove retention of its newest suffix. The CLI therefore
\emph{prepares} an O(1) anchor bundle (a Merkle root over the committed audit
chain) for an optional deployment-supplied publisher to submit to an external
witness, such as a transparency log, timestamp authority or on-chain anchor.
The released CLI does not publish periodically; nothing leaves the host unless
the deployment runs a publisher. An independently retained, published checkpoint
lets its holder detect deletion of the anchored suffix. Publication never blocks
a tool decision: witness failure reduces completeness visibility, not
enforcement. The uncovered suffix starts at the last successfully published,
independently retained checkpoint; a configured interval alone does not bound
it during publication failures.

\section{Implementation}

The TypeScript implementation provides a CLI, SDK, two runtime adapters,
local audit storage and receipt tooling. A separate backend supplies a Groth16
prover and Solidity contracts. The evaluation uses fixed published artifacts
and separately pinned unit-test and circuit companions. Each result retains its
own source identity; Appendix~\ref{sec:artifact-provenance} maps the experiments
to these artifacts and documents the verifier correction.

\subsection{Runtime adapters}

The adapters support Claude Code and OpenAI Codex CLI. Their payload handling
differs in three tested classes, described in
Section~\ref{sec:runtime-coverage}. Tests with the real Claude Code 2.1.159 and
Codex 0.155.1 binaries check hook delivery and denial by inspecting the resulting
filesystem state. These tests also record differences between synthetic adapter
inputs and the events emitted by the runtimes.

\subsection{Receipt path}

The default external verification path uses Ed25519 signatures and deterministic
audit exports. Receipt issuance first verifies the local audit chain and selects
a pre-execution decision.

\subsection{Groth16/EVM path}

The optional backend uses a versioned public-input schema and generated Solidity
verifier. Its development setup keys are not production ceremony material.
Durability unit tests and the Groth16 case study use separately pinned source
companions; they do not execute the npm CLI for their internal computations.

\section{Security analysis}
\label{sec:security-analysis}

The contract combines trusted mediation with cryptographic integrity. It is not
a proof that an untrusted host enforced policy. Table~\ref{tab:security-matrix}
separates measured properties from untested scope.

\paragraph{Mediation and persistence.}
The evaluation tests 18 synthetic adapter cells and eight real-runtime shell/file
allow/deny cases with scripted tool choices. These results support the tested
hook paths, not complete mediation of every possible runtime operation. Once a
policy is discovered, malformed input, invalid policy and tested persistence
errors block authorization. Permission-denied CLI tests and fixed-source unit
fault injection support the persistence barrier; the unit tests check file and
directory flush ordering, retries and torn-append rollback. They are not kernel
crash or power-loss tests. Neither a receipt timestamp nor a signature proves
physical execution order. The ordering claim rests on trusted implementation
control flow and this bounded instrumentation.

\paragraph{Adapter information loss.}
Commitments authenticate the post-adaptation object. They do not establish that
adaptation preserved all distinctions in the runtime request. The separate
argv probe finds equal commitments for distinct argument vectors in the
evaluated Codex adapter (Section~\ref{sec:adapter-boundary}); it is not a SHA-256 collision or a demonstrated runtime exploit.
Similarly, an absent path field can make a path rule fail to match rather than
block. These remain limitations of the evaluated implementation.

\paragraph{Evidence integrity versus consistency.}
A signature authenticates the receipt body. The export hash binds that body to
an export, but neither check establishes that a copied decision or tool name
agrees with the selected entry. That requires an explicit equality check.
Section~\ref{sec:mechanism} tests this distinction using valid research
signatures over inconsistent fields. Five omitted comparisons in the
implementation allowed such receipts to pass; adding the comparisons rejects
the identical inputs. These counterexamples concern consistency under a known
signing key, not an unprivileged Ed25519 forgery.

Other checks bind the exported action to its commitment and require the export
itself. They reject changes beyond a displayed summary and a receipt presented
without its supporting export. The failure matrix separately tests unsigned
tampering, substituted exports, wrong keys and changed policy evidence.
Together these finite tests exercise distinct verification obligations; they
do not constitute an exhaustive correctness proof.

\paragraph{Concurrency and suffix completeness.}
Twenty parallel pre-hooks produce one verified chain with all 20 entries.
A local hash chain cannot detect deletion of its newest suffix. In the separate
600-entry study, an independently retained 400-entry checkpoint detects deletion
to 350 entries, but not deletion to 550. This entry-count result does not establish
publication liveness or detection latency. External witnesses and retained
checkpoints add assumptions; the shipped CLI prepares anchors but does not run a
periodic publisher.

% Security-analysis summary matrix: each threat-model security goal restated as
% an invariant, the trusted component that enforces it, the adversarial test
% that exercises it, and whether that test is measured in this paper.
% Original evidence sources (published catp CLI 0.7.5; corrected checks and RQ4 use 0.7.6,
% source_commit 077c599 (v0.7.5), node v23.10.0, macOS x86_64):
%   RQ1a experiments/processed/adapter-conformance.json  -> tables/runtime-coverage.tex
%   RQ1b experiments/processed/runtime-in-the-loop.json  -> tables/runtime-in-the-loop.tex
%   RQ2  experiments/processed/failure-matrix.json       -> tables/failure-matrix.tex
%   RQ4  experiments/processed/receipts.json             -> tables/verification-cost.tex
%   Limitations experiments/processed/anchoring.json     -> tables/anchoring-tradeoff.tex
% Status column: \checkmark = the paired adversarial test is measured here.
% NOT measured (marked explicitly): network cost of publishing an anchor (no
% periodic publisher ships in CATP); the optional Groth16 statement family is
% narrower than the TOML policy engine and is not separately adversarially fuzzed.

\begin{table*}[t]
  \centering
  \footnotesize
  \caption{Security goals, trusted components and supporting tests.
  $\checkmark$ marks a measured adversarial test; the evidence column links to
  its results. The notes below identify untested cases, publication costs and
  the optional circuit's narrower scope.}
  \label{tab:security-matrix}
  \begin{tabular}{@{}p{2.7cm} p{3.3cm} p{6.4cm} p{2.5cm}@{}}
    \toprule
    \textbf{Invariant (security goal)} & \textbf{Trusted component} &
    \textbf{Adversarial test (measured evidence)} & \textbf{Status} \\
    \midrule
    Complete mediation within the adapter surface
    & Runtime adapter $+$ hook dispatch; the runtime honors the blocking result
    & Adapter conformance (Table~\ref{tab:runtime-coverage}): 9 payload classes
      $\times$ 2 runtimes fed to \texttt{catp hook pre}; 18/18 match, including
      fail-closed \emph{reject} on malformed events. Runtime tests
      (Table~\ref{tab:runtime-in-the-loop}): the real Claude Code and Codex
      binaries dispatch through the hook and \emph{honor} a \texttt{deny}---the
      blocked side effect is absent---8/8 across shell/file $\times$ allow/deny
    & \checkmark{} Measured \\
    \addlinespace[2pt]
    Fail-closed enforcement once a policy is present
    & Policy engine; policy discovery activates the enforcement boundary
    & Failure tests (Table~\ref{tab:failure-matrix}): non-JSON stdin, empty
      \texttt{tool\_name}, array-valued input, and an invalid policy all block
      (exit 2); benign and no-policy controls allow
    & \checkmark{} Measured \\
    \addlinespace[2pt]
    Evidence binding (action, key, policy, decision)
    & Signed receipt (Ed25519) over entry, commitment, export hash, and policy
      commitment; commitment chain
    & Failure tests: replay for another action, wrong public key, tampered payload, and
      substituted policy are all rejected; a flipped decision or altered
      commitment breaks \texttt{log verify}. Verification cost in
      Table~\ref{tab:verification-cost}
    & Five copied-field checks tested (Table~\ref{tab:mechanism}) \\
    \addlinespace[2pt]
    Audit consistency under concurrency
    & Per-agent, per-day lock-serialized append
    & Failure tests: 20 concurrent pre-hooks yield one intact chain with all 20 entries
      present and none lost
    & \checkmark{} Measured \\
    \addlinespace[2pt]
    Truncation visibility (conditional, with anchoring)
    & External witness $+$ the verifier's independent anchor copy; the local host
      stays untrusted for completeness
    & Limitations, Panel~B (Table~\ref{tab:anchoring-tradeoff}): a truncation
      below the last anchor is detected by an anchor holder, while a
      uncheckpointed suffix truncation is invisible to the local chain; the $\ddagger$ row in Table~\ref{tab:failure-matrix} reproduces the local-only case
    & \checkmark{} Detection semantics measured; publication cost \emph{not}
      measured \\
    \bottomrule
    \multicolumn{4}{@{}p{15.2cm}@{}}{\footnotesize Every test drives the published
      \texttt{catp} CLI and asserts a real exit code. Valid-signature tests isolate five missing field comparisons; the measured binding
      rows exercise action, signer key, policy commitment, and decision. Chain
      reordering and post-execution evidence issuance are subsumed by the
      prev-commitment link check and the action/export-hash binding
      respectively; neither has a dedicated adversarial row. Two quantities are
      \emph{not} measured and are flagged as such: the network cost of publishing
      an anchor (no periodic publisher ships in \systemname{}) and a dedicated
      adversarial fuzz of the optional Groth16 statement family (narrower than
      the general policy engine).} \\
  \end{tabular}
\end{table*}

\section{Evaluation}

The evaluation first tests whether verification detects disagreement between
a receipt and its supporting evidence. It then examines what adapters preserve
of runtime requests and whether the real runtimes honor the resulting decisions.
Failure tests cover policy errors, evidence tampering and concurrent appends. Timing measurements
cover the enforced hook and the default signed-receipt commands separately.
Groth16 and anchoring are evaluated off the default enforcement path.

\subsection{Experimental setup}
\label{sec:experimental-setup}

Experiments run on macOS x86\_64 (Darwin 25.6.0, build 25G83), with an Intel
Core i9-9880H at 2.30\,GHz (8 physical and 16 logical cores), 32\,GiB RAM,
an APFS filesystem, AC power and Node v23.10.0. Metadata beside the raw results
records the runtime and dependency versions, sample counts, warmups and summary
distributions. The receipt study compares identical signed inputs before and
after an implementation correction. It is a defect case study, not a designed
ablation or a comparison with another system. Performance results come from
separately pinned runs and are not used to infer a version speedup.
Appendix~\ref{sec:artifact-provenance} provides the per-experiment artifact
mapping. Scripts regenerate every numerical table from the retained results.

\subsection{Evidence consistency and failure handling}
\label{sec:failure-behavior}

\paragraph{Valid signatures over inconsistent evidence.}
\label{sec:mechanism}
A receipt can carry a valid signature over internally inconsistent evidence.
To isolate this case, a controlled comparison uses a temporary research signer
to create signatures over
receipts with one inconsistent field or evidence pairing
(Table~\ref{tab:mechanism}). This signer has the key; it does not model an
unprivileged attacker. The comparator checks only the signature and body hash
and is a research control, not a competing product. Expectations are recorded
before execution. Positive controls and exact error diagnostics distinguish
semantic rejection from unrelated command failures.

Two actions share the same displayed 200-character summary but have distinct
complete-action commitments. Changing a hidden field, rehashing the export and
re-signing the receipt still fails complete-action verification; the
signature-only control accepts. Omitting the export also leaves a valid signature
but fails full verification.

The verifier initially accepts five contradictions: signed decision, tool,
reason, matched-rule and timestamp fields that disagree with the selected entry.
It meets eight of the 13 expected outcomes. After adding the missing equality
checks, it meets 13/13 on the same receipt/export pairs, with input hashes
confirming their identity. The 29-case CLI matrix also passes after the change.
Table~\ref{tab:mechanism} reports this before/after comparison; the historical
artifacts and failed cases are retained as specified in
Appendix~\ref{sec:artifact-provenance}.

The signature-only control continues to accept inconsistent signed bodies.
This separates two checks that a verifier must perform: authenticating the body
and establishing its agreement with the evidence. Passing the former does not
imply the latter, even when the export itself is intact.
\begin{table}[t]
\centering\footnotesize
\caption{Verification of identical signed inputs before and after adding the missing copied-field checks. All signatures verify under a research key; this is not an unprivileged forgery. Artifact identities appear in Appendix~\ref{sec:artifact-provenance}.}
\label{tab:mechanism}
\begin{tabular}{llll}
\toprule
Case & Expected & \shortstack{Before\\checks} & \shortstack{After\\checks} \\
\midrule
valid control & accept & accept & accept \\
missing export & reject & reject & reject \\
other export & reject & reject & reject \\
hidden action field & reject & reject & reject \\
selected entry & reject & reject & reject \\
signed decision & reject & accept & reject \\
signed tool & reject & accept & reject \\
signed phase & reject & reject & reject \\
signed policy & reject & reject & reject \\
signed action & reject & reject & reject \\
signed timestamp & reject & accept & reject \\
signed reason & reject & accept & reject \\
signed rule & reject & accept & reject \\
\bottomrule
\end{tabular}
\end{table}

\paragraph{Failure handling and persistence.}
Table~\ref{tab:failure-matrix} reports a 29-scenario matrix using the published
CLI in isolated homes. Each row checks the actual exit code. For pre-hooks,
0 allows and 2 blocks; log verification uses 0 for intact and 1 for broken;
receipt verification uses 0 for valid and 1 for rejected. A separate durability
panel reports unit evidence (32 tests, all passing;
Table~\ref{tab:fsync-unit} highlights the 21 fault-injection and call-order checks).

The pre-hook blocks malformed input and invalid policies, allows when no policy
is present, and denies commands matched by destructive-command rules. Log
verification detects changes to retained entries or commitments. Receipt checks
reject replay for another action, a wrong verification key, a tampered payload
and substituted policy evidence. Twenty concurrent pre-hooks produce one intact
chain with no lost appends. Every CLI scenario matches its expected outcome
(29/29), and the durability panel passes 32/32. The unit results come from the
committed machine-readable fixture and are reproducible through the fixed-source
durability companion and its locked runner (evidence status:
\texttt{committed\_fixture}).

\begin{table*}[t]
  \centering
  \footnotesize
  \caption{CLI failure matrix (artifact mapping in Appendix~\ref{sec:artifact-provenance}):
  29/29 scenarios match their expectations, including positive controls.
  The $\ddagger$ row retains the local suffix-truncation limitation.
  Fault injection is reported separately in Table~\ref{tab:fsync-unit};
  valid-signature inconsistencies appear in Table~\ref{tab:mechanism}.}
  \label{tab:failure-matrix}
  \begin{tabular}{@{}p{4.7cm} p{4.5cm} p{5.2cm} c@{}}
    \toprule
    \textbf{Scenario} & \textbf{Expected} & \textbf{Observed} & \textbf{Match} \\
    \midrule
    \multicolumn{4}{@{}l}{\emph{Malformed input --- the pre-hook fails closed when a policy is present}} \\
    non-JSON stdin (policy present) & exit 2 & exit 2 & $\checkmark$ \\
    empty \texttt{tool\_name} & exit 2 & exit 2 & $\checkmark$ \\
    array-valued \texttt{tool\_input} & exit 2 & exit 2 & $\checkmark$ \\
    benign command (control) & exit 0 & exit 0 & $\checkmark$ \\
    malformed, no policy (control) & exit 0 & exit 0 & $\checkmark$ \\
    \midrule
    \multicolumn{4}{@{}l}{\emph{Policy evaluation --- an invalid policy fails closed; deny rules block}} \\
    policy missing version/rules & exit 2 & exit 2 & $\checkmark$ \\
    \texttt{rm -rf} matches deny rule & exit 2 & exit 2 & $\checkmark$ \\
    \midrule
    \multicolumn{4}{@{}l}{\emph{Evidence tampering --- the hash-chained audit log detects mutation}} \\
    untouched chain (control) & exit 0 & exit 0 & $\checkmark$ \\
    decision field flipped & exit 1 & exit 1 & $\checkmark$ \\
    stored commitment altered & exit 1 & exit 1 & $\checkmark$ \\
    newest suffix deleted$^{\ddagger}$ & exit 0 & exit 0 & $\checkmark$ \\
    \midrule
    \multicolumn{4}{@{}l}{\emph{Signed receipts --- evidence is bound to one action, one key, and one policy}} \\
    receipt vs.\ its own export (control) & exit 0 & exit 0 & $\checkmark$ \\
    receipt replayed for another action & exit 1 & exit 1 & $\checkmark$ \\
    verify with wrong public key & exit 1 & exit 1 & $\checkmark$ \\
    receipt payload tampered & exit 1 & exit 1 & $\checkmark$ \\
    receipt vs.\ true policy (control) & exit 0 & exit 0 & $\checkmark$ \\
    receipt vs.\ substituted policy & exit 1 & exit 1 & $\checkmark$ \\
    \midrule
    \multicolumn{4}{@{}l}{\emph{Concurrency --- lock-serialized chaining under 20 parallel appends}} \\
    20 concurrent pre-hooks: chain intact & exit 0 & exit 0 & $\checkmark$ \\
    20 concurrent pre-hooks: none lost & count 20 & count 20 & $\checkmark$ \\
    \midrule
    \multicolumn{4}{@{}l}{\emph{Enforcement-time binding \& durability --- CLI-observed fail-closed}} \\
    sidecar dir read-only at write & exit 2, block emitted, entries 1->1 & exit 2, block=yes, entries 1->1 & $\checkmark$ \\
    audit file read-only at append & exit 2, block emitted, entries 1->1 & exit 2, block=yes, entries 1->1 & $\checkmark$ \\
    action sidecar missing at export & log export rejects (exit != 0) & exit 1 & $\checkmark$ \\
    action sidecar altered at export & log export rejects (exit != 0) & exit 1 & $\checkmark$ \\
    sidecar renamed to 63-hex prefix & log export rejects (exit != 0) & exit 1 & $\checkmark$ \\
    valid bundle (control) & exit 0 & exit 0 & $\checkmark$ \\
    bundled action altered (offline verify) & exit 1 & exit 1 & $\checkmark$ \\
    sign with true policy (control) & exit 0 & exit 0 & $\checkmark$ \\
    sign after policy swap & receipt sign -f rejects (exit != 0) & exit 1 & $\checkmark$ \\
    verify after policy swap & exit 1 & exit 1 & $\checkmark$ \\
    \bottomrule
  \end{tabular}
\end{table*}

% RQ2 unit-level fsync / durability-order panel.
% GENERATED by gen_failure_matrix.py -- DO NOT EDIT BY HAND.
% Imported machine-readable summary: experiments/raw/failure-matrix/fsync-unit-summary.json
%   (source catp-plugin/tests/audit/durable.test.ts, catp 0.7.5, commit 077c599; runner: cd catp-plugin && node --experimental-vm-modules ../node_modules/.bin/jest --config jest.config.cjs tests/audit/durable.test.ts --json).
% These are UNIT-level in-memory fault-injection + fsync call-ORDER assertions
% against a fake filesystem, NOT OS-level CLI faults, and are kept separate from
% the CLI-observed matrix above. 32 tests, 32 passed, 0 failed
% (success=true, evidence=committed_fixture). The 21 checks below are the
% fault-injection / call-order subset highlighted for the paper.

\begin{table*}[t]
  \centering
  \footnotesize
  \caption{Unit-level durability evidence, imported from the CATP TypeScript suite
  (source pin in Appendix~\ref{sec:artifact-provenance}), separate from the CLI matrix in
  Table~\ref{tab:failure-matrix}. These 21 checks inject file-\texttt{fsync},
  directory-\texttt{fsync}, and \texttt{rename} failures and assert the syscall
  call-\emph{order} contract against an in-memory fake filesystem---32 tests total,
  all passing, imported from the committed machine-readable fixture and reproducible via its recorded runner command (evidence status: committed\_fixture). They are
  \emph{not} OS-level CLI faults; together the two layers bound the tested fault
  model without claiming arbitrary kernel-crash coverage.}
  \label{tab:fsync-unit}
  \begin{tabular}{@{}p{14.6cm} c@{}}
    \toprule
    \textbf{Unit-level durability check (in-memory fake fs)} & \textbf{Result} \\
    \midrule
    durableAppendLine (operation order) creates parents, appends line+newline, fsyncs file then parent for a new file & $\checkmark$ \\
    durableAppendLine (operation order) does not fsync the parent when appending to an existing file & $\checkmark$ \\
    durableAppendLine (operation order) propagates a file-fsync failure and still closes the descriptor & $\checkmark$ \\
    durableAppendLine (operation order) propagates a directory-fsync failure & $\checkmark$ \\
    durableAppendLine (operation order) fails instead of spinning forever when a write makes no progress & $\checkmark$ \\
    durableAppendLine (operation order) preserves the fsync error even when the cleanup close also fails & $\checkmark$ \\
    ensureDirectoryDurable creates missing components and fsyncs deepest-upward plus the existing parent & $\checkmark$ \\
    durableWriteContentAddressed writes a temp file, fsyncs it, atomically renames, then fsyncs the parent & $\checkmark$ \\
    durableWriteContentAddressed reports corruption when the existing target has different bytes & $\checkmark$ \\
    durableWriteContentAddressed propagates a rename failure, closes the fd, and removes the temp file & $\checkmark$ \\
    durableWriteContentAddressed propagates a file-fsync failure and removes the temp file & $\checkmark$ \\
    durableCreateEmptyFile creates parents, opens 0o600, fsyncs file then parent, and writes no bytes & $\checkmark$ \\
    durableCreateEmptyFile propagates a file-fsync failure and still closes the descriptor & $\checkmark$ \\
    durableAppendLine (operation order) rolls back a torn append to the original length when the write fails mid-way & $\checkmark$ \\
    durableAppendLine (operation order) rolls back after a partial write when the fsync fails, preserving both errors & $\checkmark$ \\
    durableAppendLine (operation order) truncates a torn append on a brand-new file back to empty & $\checkmark$ \\
    durableWriteContentAddressed re-establishes the durability barrier when identical bytes already exist & $\checkmark$ \\
    durableWriteContentAddressed re-runs the barrier on retry after rename succeeded but the dir fsync failed & $\checkmark$ \\
    durableWriteContentAddressed fails closed when the idempotent barrier itself cannot be flushed & $\checkmark$ \\
    durableCreateEmptyFile re-establishes the durability barrier when the file already exists & $\checkmark$ \\
    durableCreateEmptyFile re-runs the barrier on retry after the new-file dir fsync failed & $\checkmark$ \\
    \bottomrule
  \end{tabular}
\end{table*}

Successful controls include an untampered chain, a receipt checked against its
own export and a benign command. They distinguish intended rejection from an
unrelated error that causes every invocation to fail. One negative result also
matches its expectation: deleting the newest suffix leaves a self-consistent
local chain that \texttt{log verify} accepts. Detecting that deletion requires
independently retained evidence; Section~\ref{sec:limitations} tests the coverage
provided by an external checkpoint.

\subsection{Runtime coverage and adapter differences}
\label{sec:runtime-coverage}

\paragraph{Synthetic adapter inputs.}
Table~\ref{tab:runtime-coverage} maps the actions each adapter can mediate and
their differences in payload handling. Each cell feeds a hand-built PreToolUse
payload to the published \texttt{catp hook pre}. The exit code and audit append
classify the outcome as \emph{allow} (mediated, exit 0), \emph{block} (mediated
and denied, exit 2 with an entry), or \emph{reject} (the adapter returned null
and the hook failed closed, exit 2 with no entry). All 18 cells match the
expectations derived from the adapter and policy-engine source before testing.
% RQ1a adapter-conformance matrix (synthetic, runtime-shaped payloads).
% GENERATED by experiments/scripts/gen_runtime_tables.py -- DO NOT EDIT BY HAND.
% Source of record: experiments/processed/adapter-conformance.json
%   (raw per-cell results: experiments/raw/adapter-conformance/conformance.jsonl;
%    provenance: experiments/raw/adapter-conformance/meta.json -- catp 0.7.5,
%    release commit 077c599 (v0.7.5), node v23.10.0, Darwin 25.6.0 x86_64; executable sha256 6fc34688e8e5
%    cross-checked byte-identical to the published npm 0.7.5 tarball).
% Each cell drives `catp hook pre --runtime <rt>` on a HAND-BUILT payload and is
% classified from the exit code AND whether a chained audit entry was appended:
%   allow  = mediated, allowed (exit 0, +1 entry)
%   block  = mediated, denied by policy (exit 2, +1 entry)
%   reject = adapter returned null, fails closed (exit 2, +0 entries)
% Expectations were derived from src/adapters/*.ts and src/policy/engine.ts and
% all 18 cells matched (18/18). Adapter conformance ONLY: it does not show that a
% real runtime emits these payloads or honors a deny -- see RQ1b
% (rq1b_runtime_in_the_loop.sh). Regenerate: rq1_adapter_conformance.sh then
% gen_runtime_tables.py.  $\dagger$ = cross-runtime divergence in the observed class.

\begin{table}[t]
  \centering
  \footnotesize
  \caption{Adapter conformance on synthetic, runtime-shaped payloads. Each cell
  feeds a hand-built payload to \texttt{catp hook pre} and classifies it from the
  exit code and audit delta into \emph{allow}, \emph{block}, or \emph{reject};
  $\dagger$ marks the 3 rows where adapter semantics diverge. All 18 cells matched
  the expectation derived from the adapter and engine source. This suite
  establishes adapter conformance only; whether a real runtime emits such payloads
  and honors a \texttt{deny} is tested separately in
  Table~\ref{tab:runtime-in-the-loop}. Classification semantics, the three
  divergences, and the Codex fixture names are discussed in
  Section~\ref{sec:runtime-coverage}.}
  \label{tab:runtime-coverage}
  \begin{tabular}{@{}lcc@{}}
    \toprule
    \textbf{PreToolUse payload} & \textbf{Claude Code} & \textbf{Codex} \\
    \midrule
    shell, benign command & allow & allow \\
    shell, destructive (string) & block & block \\
    shell, destructive (argv array)$^{\dagger}$ & allow & block \\
    file write, benign path & allow & allow \\
    file write, path in \texttt{secrets/}$^{\dagger}$ & block & allow \\
    malformed: missing tool input & reject & reject \\
    malformed: empty tool name & reject & reject \\
    malformed: array tool input & reject & reject \\
    wrong \texttt{hook\_event\_name}$^{\dagger}$ & allow & reject \\
    \bottomrule
  \end{tabular}
\end{table}

Three differences prevent policies from transferring unchanged between runtimes.
Codex normalizes an argv-array command into a string, so the tested
\texttt{rm -rf} array is blocked. Claude Code leaves the array unchanged, and a
pattern rule that matches only string commands allows it. Claude Code emits
shell commands as strings in practice, so this synthetic gap is latent in the
shipped runtime. It nevertheless shows why the policy engine depends on the
adapter's output shape.

Path rules have a different problem. Claude Code's \texttt{Write} tool exposes
\texttt{file\_path}, allowing a \texttt{path\_denylist} to block a write to
\texttt{secrets/}. Codex's \texttt{apply\_patch} carries an opaque patch blob
without a path field, so the same rule cannot mediate it. Finally, Codex checks
\texttt{hook\_event\_name} against the phase and rejects a mismatch; Claude
Code ignores the field. Both adapters use the same enforcement and audit
pipeline, but they pass through tool names and tool-specific fields. Sharing
that pipeline does not make policies portable.

\paragraph{Argument boundaries lost during adaptation.}
\label{sec:adapter-boundary}
A separate probe imports the adapter and commitment modules from both fixed npm
installations after verifying their complete inventories. Two synthetic shell
payloads use argument vectors \texttt{[printf, \%s, "a b"]} and
\texttt{[printf, \%s, "a", "b"]}, with the same absolute executable path.
The Codex adapter joins both into the same command string, producing identical
canonical actions and commitments. Direct argv execution of
\texttt{/usr/bin/printf} produces different outputs, \texttt{a b} and
\texttt{ab}. Changing the normalized input provides a distinct-commitment
control. Both evaluated artifacts exhibit this behavior; the accompanying
material contains the script and observations.

The distinction is lost before hashing, so this is not a cryptographic collision.
The probe uses synthetic inputs and issues no receipts. It establishes neither
reachability through a current Codex runtime nor an authorization exploit.
Full-action binding covers the adapter-normalized object. Additional receipt
checks cannot resolve this loss of argument boundaries.

\paragraph{Hook delivery and denial in the real runtimes.}
Table~\ref{tab:runtime-in-the-loop} tests the real Claude Code 2.1.159 and Codex
0.155.1 binaries with an $8$-row matrix (runtime $\times$ shell/file $\times$
allow/deny). For each attempted tool call, the harness checks hook delivery,
adapter parsing, CATP's decision and the runtime's response. It then inspects the
target filesystem state, verifies the audit chain with \texttt{catp log verify},
and checks the pre-entry's v4 policy and action commitments and the presence of
the content-addressed action sidecar. All $8/8$ rows satisfy all eight checks.
Every denial suppresses the action: the target is absent, the hook exits with
code~2, and no post-execution entry appears. Every allowed action runs and is
audited.

\begin{table*}[t]
  \centering
  \footnotesize
  \caption{Pre-execution mediation in Claude Code 2.1.159 and Codex
  0.155.1. A stub selects the tool call; the runtime and hook are real,
  with no cloud account or network egress. Each row dispatches a
  \texttt{PreToolUse} hook to the pinned CATP artifact and records the eight
  checks listed below ($\checkmark$ denotes success). All 8/8
  rows pass: denied actions leave no target (exit~2), while allowed actions
  execute and are audited. Codex file edits use an \texttt{apply\_patch} heredoc
  inside \texttt{exec\_command}, reported as \texttt{tool\_name=Bash}; Claude
  Code uses its native \texttt{Write} tool.}
  \label{tab:runtime-in-the-loop}
  \begin{tabular}{@{}ll ll ccc l cc cc@{}}
    \toprule
    \textbf{Runtime} & \textbf{Action} & \textbf{Policy} & \textbf{tool\_name}
      & \textbf{1} & \textbf{2} & \textbf{3} & \textbf{4: decision}
      & \textbf{5} & \textbf{6: side effect} & \textbf{7} & \textbf{8: v4 binding} \\
    \midrule
    Claude Code & file & allow & \texttt{Write} & $\checkmark$ & $\checkmark$ & $\checkmark$ & allow & $\checkmark$ & created & $\checkmark$ & $\checkmark$ \\
    Claude Code & file & deny & \texttt{Write} & $\checkmark$ & $\checkmark$ & $\checkmark$ & deny & $\checkmark$ & absent & $\checkmark$ & $\checkmark$ \\
    Claude Code & shell & allow & \texttt{Bash} & $\checkmark$ & $\checkmark$ & $\checkmark$ & allow & $\checkmark$ & created & $\checkmark$ & $\checkmark$ \\
    Claude Code & shell & deny & \texttt{Bash} & $\checkmark$ & $\checkmark$ & $\checkmark$ & deny & $\checkmark$ & absent & $\checkmark$ & $\checkmark$ \\
    \midrule
    Codex & file (bash-mediated) & allow & \texttt{Bash} & $\checkmark$ & $\checkmark$ & $\checkmark$ & allow & $\checkmark$ & created & $\checkmark$ & $\checkmark$ \\
    Codex & file (bash-mediated) & deny & \texttt{Bash} & $\checkmark$ & $\checkmark$ & $\checkmark$ & deny & $\checkmark$ & absent & $\checkmark$ & $\checkmark$ \\
    Codex & shell & allow & \texttt{Bash} & $\checkmark$ & $\checkmark$ & $\checkmark$ & allow & $\checkmark$ & created & $\checkmark$ & $\checkmark$ \\
    Codex & shell & deny & \texttt{Bash} & $\checkmark$ & $\checkmark$ & $\checkmark$ & deny & $\checkmark$ & absent & $\checkmark$ & $\checkmark$ \\
    \bottomrule
  \end{tabular}
  \par\vspace{3pt}
  {\footnotesize Eight checks: \textbf{1}~stub emitted the tool call;
  \textbf{2}~runtime invoked the hook (raw payload saved); \textbf{3}~the adapter
  parsed it into a valid action (pre-audit entry); \textbf{4}~\systemname{}'s
  decision; \textbf{5}~the runtime honored it; \textbf{6}~ground-truth filesystem
  terminal state of the target; \textbf{7}~entry written and \texttt{catp log
  verify} reports the chain intact; \textbf{8}~pre-entry is a valid v4 record
  binding the enforcement-time policy and action commitments, with the
  content-addressed action sidecar present (required on allow rows).}
\end{table*}

A local deterministic endpoint chooses a fixed tool call in place of a model,
without a cloud account, subscription or network egress. The runtime binary,
hook dispatch, CATP's parsing and decision, audit writes and filesystem effects
are real. This setup isolates enforcement behavior from tool selection.

The tests also exposed an integration constraint. A logging wrapper that captures
hook stdout and replays it after exit breaks Codex's denial handling: Codex reads
the live pipe concurrently and misses the late write. Forwarding bytes as they
arrive, as a \texttt{tee}-style wrapper does, preserves enforcement and evidence.

This Codex build reports \texttt{tool\_name} \texttt{Bash} for both shell commands
and file edits because it routes \texttt{apply\_patch} through
\texttt{exec\_command}. The synthetic fixture names
\texttt{shell}/\texttt{apply\_patch} are therefore adapter conventions, not the
exact names emitted by the runtime. In the one allowed Codex file-edit case,
no \texttt{PostToolUse} event appears, although one appears for a plain shell
command. That row's evidence consists of the pre-execution entry and intact
chain. The missing post-event is a runtime dispatch behavior; pre-execution
mediation still occurs.

\subsection{Enforcement overhead}
\label{sec:enforcement-cost}

Table~\ref{tab:overhead} measures the wall time of the \texttt{catp hook pre}
subprocess on both adapters. The reference is the same CLI invocation with no
policy present. It includes process start, Node/CLI initialization and hook
dispatch, with medians of 144.6\,ms for Claude Code and 141.6\,ms for Codex.
This is a reference for this CLI on this testbed, not a general process-spawn cost.

Enforcement adds median paired deltas of 85.19\,ms and 85.43\,ms for Claude Code
and Codex, respectively. Mean paired deltas are 88.04\,ms and 87.22\,ms; every
percentile-bootstrap 95\% interval excludes zero. The median increase is roughly
60\% of the no-policy baseline. Sustained mediated throughput is 4.26 and
4.33 calls/s, respectively.
% RQ3 pre-execution mediation overhead (both runtimes, randomized A/B order,
% percentile-bootstrap CI, throughput).
% GENERATED by experiments/scripts/gen_cost_tables.py -- DO NOT EDIT BY HAND.
% Source of record: experiments/processed/overhead.json
%   (raw samples: experiments/raw/overhead/latency.jsonl; provenance:
%    experiments/raw/overhead/meta.json -- catp 0.7.5, release commit
%    077c599 (v0.7.5), node v23.10.0, Darwin 25.6.0 x86\_64; executable sha256 6fc34688e8e5
%    cross-checked byte-identical to the published npm 0.7.5 tarball).
% Testbed: Darwin 25.6.0 x86\_64 (build 25G83), Intel(R) Core(TM) i9-9880H CPU @ 2.30GHz (8 physical / 16 logical cores), 32\,GiB RAM, apfs filesystem, AC power, Node v23.10.0, \systemname{} 0.7.5 (commit 077c599)
% 200 paired samples/runtime after 10 warm-up (caches warm:
% yes); seed 20260923; 10{,}000 bootstrap iterations.
% "Baseline" is the CATP Node CLI no-policy baseline: the same CLI and process
% start with no policy present, so the pre-hook exits 0 without load/evaluate/
% append. It is NOT a universal process-spawn floor and is NOT generalized to
% other external-command mediators. Net delta = paired per-iteration (enforced -
% baseline); the enforced path exercises the durable action-sidecar write +
% chained audit append + fsync. Regenerate: rq3_overhead.sh then
% gen_cost_tables.py.

\begin{table*}[t]
  \centering
  \footnotesize
  \caption{Wall time of \texttt{catp hook pre} on both runtime adapters.
  The baseline invokes the same CLI with no policy present, including process
  start and initialization; it is specific to this CLI and testbed.
  Enforcement adds policy loading and evaluation, durable action-sidecar
  storage and a hash-chained audit append with \texttt{fsync}. The paired
  delta measures these operations together. Testbed, sampling and bootstrap
  details appear below the table.}
  \label{tab:overhead}
  \begin{tabular}{@{}l r r r l r@{}}
    \toprule
    \textbf{Runtime} & \textbf{Baseline p50} & \textbf{Enforced p50}
      & \textbf{Net $\Delta$ p50} & \textbf{Net $\Delta$ mean [95\% boot CI]}
      & \textbf{Throughput} \\
    & \textbf{(ms)} & \textbf{(ms)} & \textbf{(ms)} & \textbf{(ms)}
      & \textbf{(calls/s)} \\
    \midrule
    Claude Code & 144.6 & 230.4 & 85.19 & 88.04\,[85.44,\,91.22] & 4.26 \\
    Codex & 141.6 & 227.4 & 85.43 & 87.22\,[85.27,\,89.17] & 4.33 \\
    \bottomrule
  \end{tabular}
  \par\vspace{3pt}
  {\footnotesize Testbed: Section~\ref{sec:experimental-setup}; artifacts: Appendix~\ref{sec:artifact-provenance}. 200 paired samples per runtime after 10
  warm-up (caches warm), randomized A/B order (seed 20260923), 10{,}000 bootstrap
  iterations; every bootstrap interval on the mean delta excludes zero, and the median net delta is 85.19\,ms (Claude Code)
  / 85.43\,ms (Codex), for the two tested adapters.}
\end{table*}

Each iteration pairs an enforced call with a no-policy call under the same
workload. Their order is randomized with seed 20260923. A single driver process
times 200 pairs per runtime after 10 warm-up iterations, excluding driver startup
from each sample and reducing sensitivity to slow thermal drift. Mean-delta
intervals use a distribution-free percentile bootstrap with 10{,}000 resamples,
without assuming symmetric latency.

The paired delta covers policy loading and evaluation, creation of a real
content-addressed action sidecar (temporary write, file \texttt{fsync}, rename
and directory \texttt{fsync}), and the chained audit append with its own
\texttt{fsync}. The experiment measures these together on every timed sample.
Isolating their costs or attributing changes to a release would require matched
component ablations or a fixed-workload version comparison. Neither was run.
No model-based mediator was benchmarked, so the measurements also provide no
latency comparison with semantic classification.

\subsection{Receipt issuance and offline verification}

Table~\ref{tab:verification-cost} reports the default off-chain signed-receipt
backend. Issuance re-verifies the audit chain, exports the entry and signs it
with Ed25519. Its median is 183.6\,ms (mean $187.4 \pm 5.1$\,ms, $n{=}100$).
Verification against the audit export and policy commitment takes a median
174.4\,ms (mean $177.2 \pm 3.2$\,ms, $n{=}100$). These are end-to-end wall times
of \texttt{catp receipt issue} and \texttt{catp receipt verify} on the same
testbed after 10 warmups, using the verifier with all copied-field checks
(Appendix~\ref{sec:artifact-provenance}).

The receipt measurements use different subcommands from
\texttt{hook pre}, so the no-policy hook reference is not subtracted.
Verification rehashes the export with SHA-256, checks an Ed25519 signature and
compares the policy commitment. It is read-only, with no durable append or
directory \texttt{fsync}. A receipt is 977\,bytes and its audit-export bundle is
978\,bytes. The verifier's only independently trusted state is the 113-byte
Ed25519 public key. Offline verification also needs the receipt, export and
canonical action, plus policy evidence when that commitment is checked.
% RQ4 verification cost: the DEFAULT signed-receipt backend (off-chain).
% GENERATED by experiments/scripts/gen_cost_tables.py -- DO NOT EDIT BY HAND.
% The optional Groth16/EVM backend is presented separately as a case study in
% tables/groth16-case-study.tex (it is opt-in, not the default path).
% Source of record: experiments/remediation-0.7.6/run-20260927-01/processed/receipts.json  ("receipts" block)
%   (raw receipt timings: experiments/remediation-0.7.6/run-20260927-01/raw/receipts/receipts.jsonl; provenance:
%    experiments/remediation-0.7.6/run-20260927-01/raw/receipts/meta.json -- catp 0.7.6, release commit
%    fec7e8e (v0.7.6), node v23.10.0, Darwin 25.6.0 x86\_64; executable sha256 6fc34688e8e5).
% Testbed: Darwin 25.6.0 x86\_64 (build 25G83), Intel(R) Core(TM) i9-9880H CPU @ 2.30GHz (8 physical / 16 logical cores), 32\,GiB RAM, apfs filesystem, AC power, Node v23.10.0, \systemname{} 0.7.6 (commit fec7e8e)
% All values MEASURED by this run ($n{=}100$ after 10 warm-up); CI is
% +/- 1.96*SEM. Regenerate: run_remediation.py NEW_RUN --select then gen_cost_tables.py.

\begin{table*}[t]
  \centering
  \footnotesize
  \caption{Verification cost of the \emph{default} signed-receipt backend
  (off-chain). Issuing a receipt re-verifies the audit chain, exports the entry,
  and signs it with Ed25519; verifying checks the signature against the audit
  export and the policy commitment. Both are end-to-end wall times of the
  \systemname{} Node CLI receipt commands. These are absolute times: the
  no-policy hook is a different workload and is not subtracted. No component
  or crypto-only timing is inferred. A receipt is
  977\,bytes; the 113\,bytes of public-key material is
  the only independently trusted state, while offline verification also consumes
  the receipt, the audit export, and the canonical action. Testbed: Section~\ref{sec:experimental-setup}; artifacts: Appendix~\ref{sec:artifact-provenance};
  $n{=}100$ after 10 warm-up. The optional on-chain Groth16 backend is
  quantified separately in Table~\ref{tab:groth16-case-study}.}
  \label{tab:verification-cost}
  \begin{tabular}{@{}p{5.5cm} r p{6.5cm}@{}}
    \toprule
    \textbf{Metric} & \textbf{Value} & \textbf{Detail} \\
    \midrule
    Issue latency (median / mean)   & 183.6 / 187.4\,ms & measured, $n{=}100$, CI $\pm5.1$\,ms \\
    Verify latency (median / mean)  & 174.4 / 177.2\,ms & measured, $n{=}100$, CI $\pm3.2$\,ms \\
    Receipt size                    & 977\,B            & measured \\
    Audit-export bundle size        & 978\,B            & measured \\
    Signature / public key          & Ed25519 / 113\,B  & measured \\
    \bottomrule
  \end{tabular}
\end{table*}

\subsection{The optional Groth16/EVM backend}
\label{sec:groth16-case-study}

The optional backend supports EVM verification of a narrower circuit-defined
authorization predicate. It does not implement the full local policy language
or receipt/export contract and is separate from default enforcement.
Table~\ref{tab:groth16-case-study} reports an average proving step of 50.8\,ms
for the 13{,}284-constraint bn254 circuit and native verification of 0.86\,ms,
under 1\,ms. A one-shot invocation includes circuit compilation and constraint
construction, giving a median end-to-end proving time of 503.4\,ms.

On Sepolia, executing an authorized action consumed 351{,}706 gas and registering
a policy consumed 91{,}253 gas. The verifier runtime is 6{,}365 bytes and the
authorization wrapper is 1{,}063 bytes. These figures come from committed
deployment metadata (transaction \texttt{0x6e45\ldots d39}, chain id 11155111,
deployed 2026-08-31, status \texttt{smoke\_passed}); only proving costs were
measured locally ($n{=}5$). Gas and milliseconds measure different resources,
so no ratio is computed between them. The default signed receipt requires no
on-chain transaction.
% Case study: cost of the OPTIONAL Groth16/EVM alternative verifier for the same
% authorization. This is NOT the default verification path (signed receipts are);
% it is a case study of what a deployment pays for smart-contract-enforceable,
% publicly verifiable authorization -- an alternative verifier, not a core
% protocol dependency.
% GENERATED by experiments/scripts/gen_cost_tables.py -- DO NOT EDIT BY HAND.
% Source of record: experiments/processed/receipts.json  ("groth16" block)
%   (raw: experiments/raw/receipts/groth16.json; provenance:
%    experiments/raw/receipts/meta.json -- catp 0.7.5, release commit
%    077c599 (v0.7.5), node v23.10.0, Darwin 25.6.0 x86\_64).
% Inputs are consumed ONLY from the pinned companion artifact
%   experiments/artifact/companion/groth16 (git archive of tag v0.7.5, verified
%   against its SHA256SUMS before use; subtree sha256 3b5b2f9f6009) -- never a live
%   CATP checkout. Provenance column distinguishes numbers MEASURED locally from
%   numbers CAPTURED from the companion's committed artifacts (never re-derived):
%     measured ($n{=}5$)  -> prove wall time + gnark prover/local-verify steps
%     circuit manifest         -> keys/*.manifest.json
%     Sepolia deploy metadata  -> deployment.sepolia-groth16.json
% METRIC DISCIPLINE: proving time and native verification time are wall-clock
%   MILLISECONDS; on-chain verification is EVM GAS. They are DISTINCT,
%   non-comparable metrics; no cross-unit ratio is computed or claimed.
% Regenerate: rq4_verification_cost.sh then gen_cost_tables.py.

\begin{table*}[t]
  \centering
  \footnotesize
  \caption{Cost of the optional Groth16/EVM backend for its narrower
  circuit-defined authorization predicate. It does not implement the full local
  policy or receipt/export contract; signed receipts remain the default
  (Table~\ref{tab:verification-cost}). Local proving and native verification
  times are measured in milliseconds; on-chain costs are gas from the committed
  Sepolia deployment. These units do not support a direct cost ratio.
  The gnark prover step averages 50.8\,ms for the
  13{,}284-constraint bn254 circuit, with native verification of
  0.86\,ms. Including compilation and constraint construction gives
  a median end-to-end proving time of 503.4\,ms. Authorized execution
  consumed 351{,}706 gas and policy registration 91{,}253 gas, with a
  6{,}365-byte verifier and 1{,}063-byte wrapper.
  Gas and bytecode sizes come from committed Sepolia metadata
  (tx \texttt{0x6e45\ldots d39}); proving costs were measured locally ($n{=}5$).
  The pinned source companion (Appendix~\ref{sec:artifact-provenance}) is verified against its SHA256SUMS before use.}
  \label{tab:groth16-case-study}
  \begin{tabular}{@{}p{5.5cm} r p{6.5cm}@{}}
    \toprule
    \textbf{Metric} & \textbf{Value} & \textbf{Provenance} \\
    \midrule
    gnark prover step (mean) & 50.8\,ms & measured, $n{=}5$ \\
    gnark local verify step (mean) & 0.86\,ms & measured, $n{=}5$ \\
    Prove end-to-end (median / mean) & 503.4 / 718.7\,ms & measured, incl.\ one-shot circuit compile \\
    Proof artifact (JSON) & 2{,}035\,B & measured \\
    Proof size (canonical) & 256\,B & circuit manifest (bn254, MiMC) \\
    Constraints / public inputs & 13{,}284 / 13 & circuit manifest \\
    \midrule
    On-chain executeAuthorized gas & 351{,}706 & Sepolia deploy (tx \texttt{0x6e45\ldots d39}) \\
    On-chain registerPolicy gas & 91{,}253 & Sepolia deploy \\
    Verifier deploy / total gas & 1{,}428{,}983 / 2{,}460{,}585 & Sepolia deploy \\
    Verifier / wrapper runtime & 6{,}365 / 1{,}063\,B & Sepolia deploy (EIP-170 limit 24{,}576) \\
    \bottomrule
  \end{tabular}
\end{table*}

\subsection{Checkpoint coverage and remaining limits}
\label{sec:limitations}

A local hash chain cannot reveal deletion of its newest suffix, as the
$\ddagger$ row of Table~\ref{tab:failure-matrix} demonstrates.
Table~\ref{tab:anchoring-tradeoff} measures anchor preparation and tests what an
independently retained checkpoint adds. Panel~(A) times \texttt{catp anchor},
which re-verifies the chain and computes a Merkle root over all commitments.
Preparation scales linearly at $20.3\,\mu$s per entry above a fitted intercept
of approximately 180\,ms; the intercept's components are not isolated. Anchor
preparation runs off the enforcement path and is not included in per-call
mediation. The witness-facing bundle is O(1), at 191--192 bytes; only the digit
width of the entry count depends on chain length.

Panel~(B) starts with 600 entries and retained anchors at 200 and 400
(interval $T{=}200$). Both a 600$\to$550 cut and a 600$\to$350 cut pass local
\texttt{log verify}. An auditor holding the 400-entry anchor detects the latter
because 350 entries cannot reproduce that prefix. Deletion confined to entries
after the checkpoint remains invisible to both checks; the uncheckpointed suffix
contains 200 entries. A configured interval bounds coverage only if publication,
independent checkpoint retention and verifier checks succeed. Outages can extend
the uncovered suffix, so these entry-count results do not establish a wall-clock
detection bound.

\begin{table*}[t]
  \centering
  \footnotesize
  \caption{Anchor preparation and checkpoint coverage. The released CLI
  prepares an O(1) bundle; publication requires a deployment-supplied publisher
  and an external witness. Publication cost was not measured.
  \textbf{(A)} Preparation runs off the enforcement path and scales linearly
  with chain length; the \texttt{catp\_audit\_anchor\_v1} bundle stays O(1).
  \textbf{(B)} On a 600-entry chain with retained anchors at 200 and
  400 ($T{=}200$ entries), \texttt{log verify} detects a rewritten
  entry but accepts both tested tail truncations. An auditor holding the last
  anchor also detects truncation below that checkpoint. The uncheckpointed
  suffix in this fixture contains $T$ entries; no wall-clock detection bound
  is measured. All 4 cases match their expectations. Testbed and fit
  provenance appear below the tables.}
  \label{tab:anchoring-tradeoff}

  \begin{tabular}{@{}r r r r@{}}
    \toprule
    \multicolumn{4}{@{}l}{\emph{(A) Anchor-preparation cost vs.\ chain length
      (\texttt{catp anchor}, $n{=}15$ per point)}} \\
    \addlinespace[2pt]
    \textbf{Entries $N$} & \textbf{p50 (ms)} & \textbf{mean $\pm$ CI95 (ms)} & \textbf{Bundle (B)} \\
    \midrule
    1{,}000   & 169.2 & 169.5 $\pm$ 1.5 & 191 \\
    5{,}000   & 260.9 & 261.5 $\pm$ 2.1 & 191 \\
    10{,}000   & 366.1 & 367.4 $\pm$ 6.6 & 192 \\
    25{,}000   & 803.9 & 811.5 $\pm$ 23.5 & 192 \\
    50{,}000   & 1138.5 & 1139.2 $\pm$ 18.9 & 192 \\
    \addlinespace[1pt]
    \multicolumn{4}{@{}l}{\footnotesize fit: $20.3\,\mu$s/entry $+\,180.5$\,ms
      intercept; bundle O(1) in $N$} \\
    \bottomrule
  \end{tabular}

  \vspace{6pt}

  \begin{tabular}{@{}l c l l@{}}
    \toprule
    \multicolumn{4}{@{}l}{\emph{(B) Truncation detection on a 600-entry chain,
      anchors held at 200 and 400 ($T{=}200$)}} \\
    \addlinespace[2pt]
    \textbf{Case} & \textbf{Shown} & \textbf{Local \texttt{log verify}} & \textbf{Held external anchor} \\
    \midrule
    Intact chain (control) & 600 & accept & accept (root match) \\
    Truncate tail, inside window & 550 & accept & accept \emph{(blind spot)} \\
    Truncate below last anchor & 350 & accept & \textbf{detect} (count short) \\
    Rewrite inside anchored prefix & 600$^{*}$ & \textbf{detect} (broken) & \textbf{detect} (broken) \\
    \bottomrule
    \multicolumn{4}{@{}l}{\footnotesize $^{*}$entry 100 mutated. Undetectable
      window $= N_{\text{true}} - N_{\text{last anchor}} = 200 = T$.} \\
  \end{tabular}
  \par\vspace{3pt}
  {\footnotesize Testbed: Section~\ref{sec:experimental-setup}; artifacts: Appendix~\ref{sec:artifact-provenance}. Panel~(A) fit: least squares over
  $N{=}1000\ldots50000$, $n{=}15$ per point; bundle O(1) in $N$.}
\end{table*}

The released CLI prepares anchors but has no periodic publisher. Publication
cost depends on the chosen transparency log, timestamp authority or chain and
was not measured. The O(1) bundle size is the only measured witness-facing
quantity; neither a publication schedule nor detection latency follows from it.

The remaining limits concern what the tests and policy language can establish.
First-match command globs do not parse shell semantics, unmatched tools are
allowed, and path rules do not resolve symlinks. Adapter normalization can hide
path structure even when hook delivery works. Runtime mediation still depends
on hook delivery and denial handling, and the signer authenticates assertions
rather than host honesty or execution outcomes. Finite consistency tests do not
prove exhaustive correctness, and scripted tool choices do not measure autonomous
task utility. The durability evidence covers mocked filesystem faults and CLI
permission errors, not process kills, kernel crashes or power loss.

\section{Related work}
\label{sec:related-work}

We compare systems by the boundary they enforce, the statement their evidence
supports, and the parties a verifier must trust. Authorization systems can
produce cryptographic evidence, and audit systems can
bind policies to actions, so these categories overlap. The comparisons below
use documented mechanisms; an undocumented feature is not evidence of its absence.

\subsection{Agent authorization and runtime enforcement}

Progent checks symbolic rules over tool names and arguments and restricts
unapproved policy changes to privilege narrowing~\cite{shi2025progent}.
AgentBound enforces declarative permissions at MCP execution
boundaries~\cite{buhler2026agentbound}. AgentGuard studies attribute-based
authorization for tool-use agents~\cite{agentguard2026}, and AgentSpec provides
customizable runtime rules~\cite{agentspec2025}. These works inform the policy
and mediation side of our design. CATP does not claim a more expressive policy
language or better attack prevention; it studies how the resulting decision is
bound to evidence that survives outside the runtime.

CaMeL separates control flow from untrusted data~\cite{debenedetti2025camel},
IsolateGPT studies execution isolation~\cite{isolategpt}, ACE separates
authorization from planning~\cite{ace2025}, and SEAgent applies mandatory access
control to privilege escalation~\cite{seagent2026}. These address boundaries
that a signed receipt alone cannot secure. SAGA is especially relevant because
it combines agent governance with cryptographically derived access-control
tokens and formal guarantees~\cite{syros2026saga}. Its inter-agent authorization
and delegation focus differs from CATP's local hook, durable decision record,
and later receipt export. The distinction is the protected boundary and
verified statement, not the presence versus absence of cryptography.

\subsection{Accountability and verifiable records}

Classical protection principles include complete mediation and fail-safe
design~\cite{saltzer1975protection};
CATP relies on runtime hooks rather than establishing an OS reference monitor.
Signed logs are a longstanding accountability technique. PeerReview uses
tamper-evident records to detect deviations from specified behavior in
distributed systems~\cite{haeberlen2007peerreview}. CATP neither introduces
signed logging nor inherits PeerReview's accountability guarantees merely by
using a hash chain.

Agent Flight Recorder records structured agent events and evaluates external
anchoring strategies~\cite{agentflightrecorder2026}. Its policy and intent fields
overlap our evidence schema; our emphasis is the ordering and binding between a
local authorization decision and durable evidence. Notarized Agents (Sello)
instead places attestation at the receiving service and uses a witness-cosigned
log~\cite{notarizedagents2026}. That trust boundary can support claims about
what a receiver observed, whereas CATP's host-side receipt authenticates a
monitor's decision and does not independently establish execution. AuditableLLM
studies hash-chain-backed compliance auditing~\cite{auditablellm2026}.

The SCITT architecture separates signed statements from transparency receipts
for their registration~\cite{scitt2025architecture}. Registration evidence does
not by itself establish that a local runtime enforced a decision. Conversely,
CATP's local chain does not provide an independently witnessed complete history:
a verifier needs retained external evidence to detect deletion of checkpointed
records, and uncheckpointed events remain outside that guarantee.

\subsection{Public implementations and emerging profiles}

Signet's public documentation describes policy attestations, intent-bound
authorization decisions, delegation, and offline-verifiable
receipts~\cite{signet2026}. These mechanisms directly overlap CATP's design.
The source is software documentation; we have not audited the implementation or run a comparative benchmark.

The Permit profile specifies pre-execution authorization records binding a
canonical request, a decision, and verification material, with dispatch checking
the request digest~\cite{munoz2026permit}. This overlaps CATP's central problem.
It is an individual Internet-Draft, not an adopted IETF standard or a
peer-reviewed evaluation. The comparison concerns documented mechanisms; security and performance have
not been measured across these systems. Keel documents managed pre-execution decisions and signed
request-bound evidence, with route-specific guarantees~\cite{keel2026}. This
is another direct implementation neighbor; we have not independently verified
its claimed coverage or benchmarked it.

\subsection{Scope of the evaluation}

Prompt-injection benchmarks and defenses, including AgentDojo and SecAlign,
evaluate attack and task outcomes~\cite{debenedetti2024agentdojo,secalign2025}.
Our deterministic runtime matrix tests hook delivery, denial, side effects,
and evidence integrity; it does not measure autonomous task utility or
prompt-injection robustness. Similarly, an optional Groth16 circuit does not
establish a new proof construction or equivalence to the full local policy
language.

CATP's evaluation follows the decision from the runtime hook through durable
storage to an offline receipt check. It tests whether the implementation binds
the policy and normalized action used at decision time, persists evidence before
permission and checks the export against the receipt. Related systems could
enforce the same constraints. The results here document CATP's implementation of these constraints,
observed failures and end-to-end costs in two runtime integrations.

\section{Conclusion}

Portable authorization evidence requires agreement between the runtime decision,
the durable record and the signed receipt. CATP specifies these bindings and
implements them at the pre-execution hooks of two agent runtimes. Validly signed
counterexamples demonstrate why authenticating a receipt body is insufficient:
its copied fields must also agree with the selected audit entry. Explicit
comparisons reject contradictions that otherwise pass signature and export
checks. The evaluated verifier satisfies all 13 expectations after adding the
missing comparisons.

The resulting evidence still refers to the adapter's representation of an
action. The argument-vector probe shows that this representation can discard a
meaningful distinction before hashing. Verification cannot recover it, and
policy portability depends on the fields each adapter exposes. Trust in the
runtime, host and signer therefore remains necessary; a consistent receipt
alone proves neither safe execution nor complete disclosure. Retained
checkpoints cover earlier prefixes, while the optional Groth16 backend checks
only its circuit-defined predicate.

The eight scripted runtime cases confirm hook delivery and the tested
allow and deny behavior.
Median paired hook overhead is approximately 85\,ms and offline receipt
verification takes approximately 174\,ms on the testbed. These full-path
measurements make the implementation cost explicit, without attributing it to
individual components or implying a version speedup. Autonomous task utility
and comparative effectiveness against other systems remain unmeasured.

% Bibliography enabled: all cited entries verified against primary sources
% (see bibliography/references-audit.md for dates and publication status).
\bibliographystyle{ACM-Reference-Format}
\bibliography{bibliography/references}

@inproceedings{syros2026saga,
  title     = {SAGA: A Security Architecture for Governing AI Agentic Systems},
  author    = {Syros, Georgios and Suri, Anshuman and Ginesin, Jacob and
               Nita-Rotaru, Cristina and Oprea, Alina},
  booktitle = {Proceedings of the Network and Distributed System Security
               Symposium (NDSS)},
  year      = {2026}
}

@misc{debenedetti2025camel,
  title         = {Defeating Prompt Injections by Design},
  author        = {Debenedetti, Edoardo and Shumailov, Ilia and Fan, Tianqi and
                   Hayes, Jamie and Carlini, Nicholas and Fabian, Daniel and
                   Kern, Christoph and Shi, Chongyang and Terzis, Andreas and
                   Tram{\`e}r, Florian},
  year          = {2025},
  eprint        = {2503.18813},
  archiveprefix = {arXiv},
  primaryclass  = {cs.CR}
}

@misc{agentguard2026,
  title         = {AgentGuard: An Attribute-Based Access Control Framework for
                   Tool-Use LLM-Based Agent},
  author        = {Luo, Jiaqi and Peng, Songyang and Dai, Jiarun and
                   Chen, Zhile and Shen, Zhuoxiang and Hong, Geng and
                   Pan, Xudong and Zhang, Yuan and Yang, Min},
  year          = {2026},
  eprint        = {2605.28071},
  archiveprefix = {arXiv},
  primaryclass  = {cs.CR}
}

@misc{shi2025progent,
  title         = {Progent: Securing AI Agents with Privilege Control},
  author        = {Shi, Tianneng and He, Jingxuan and Wang, Zhun and
                   Li, Hongwei and Wu, Linyu and Guo, Wenbo and Song, Dawn},
  year          = {2025},
  eprint        = {2504.11703},
  archiveprefix = {arXiv},
  primaryclass  = {cs.CR}
}

@inproceedings{buhler2026agentbound,
  title     = {AgentBound: Securing Execution Boundaries of AI Agents},
  author    = {B{\"u}hler, Christoph and Biagiola, Matteo and
               Di Grazia, Luca and Salvaneschi, Guido},
  booktitle = {Proceedings of the 34th ACM Joint European Software Engineering
               Conference and Symposium on the Foundations of Software
               Engineering (FSE)},
  year      = {2026},
  doi       = {10.1145/3808103}
}

@inproceedings{agentspec2025,
  title     = {AgentSpec: Customizable Runtime Enforcement for Safe and
               Reliable LLM Agents},
  author    = {Wang, Haoyu and Poskitt, Christopher M. and Sun, Jun},
  booktitle = {Proceedings of the 48th IEEE/ACM International Conference on
               Software Engineering (ICSE)},
  year      = {2026}
}

@inproceedings{isolategpt,
  title     = {IsolateGPT: An Execution Isolation Architecture for LLM-Based
               Agentic Systems},
  author    = {Wu, Yuhao and Roesner, Franziska and Kohno, Tadayoshi and
               Zhang, Ning and Iqbal, Umar},
  booktitle = {Proceedings of the Network and Distributed System Security
               Symposium (NDSS)},
  year      = {2025},
  doi       = {10.14722/ndss.2025.241131}
}

@inproceedings{ace2025,
  title     = {ACE: A Security Architecture for LLM-Integrated App Systems},
  author    = {Li, Evan and Mallick, Tushin and Rose, Evan and
               Robertson, William and Oprea, Alina and Nita-Rotaru, Cristina},
  booktitle = {Proceedings of the Network and Distributed System Security
               Symposium (NDSS)},
  year      = {2026},
  doi       = {10.14722/ndss.2026.230352}
}

@misc{seagent2026,
  title         = {Taming Various Privilege Escalation in LLM-Based Agent
                   Systems: A Mandatory Access Control Framework},
  author        = {Ji, Zimo and Wu, Daoyuan and Jiang, Wenyuan and
                   Ma, Pingchuan and Li, Zongjie and Gao, Yudong and
                   Wang, Shuai and Li, Yingjiu},
  year          = {2026},
  eprint        = {2601.11893},
  archiveprefix = {arXiv},
  primaryclass  = {cs.CR}
}

@misc{agentflightrecorder2026,
  title     = {Agent Flight Recorder: Tamper-Evident Audit Trails with On-Chain
               Anchoring for Long-Horizon Tool-Using Agents},
  author    = {Bindschaedler, Laurent and Botha, Quentin and
               Siebenbrunner, Christoph},
  eprint = {2609.01931},
  archiveprefix = {arXiv},
  year      = {2026}
}

@misc{notarizedagents2026,
  title         = {Notarized Agents: Receiver-Attested Confidential Receipts for
                   AI Agent Actions},
  author        = {Figuera, Juan},
  year          = {2026},
  eprint        = {2606.04193},
  archiveprefix = {arXiv},
  primaryclass  = {cs.CR}
}

@article{auditablellm2026,
  title   = {AuditableLLM: A Hash-Chain-Backed, Compliance-Aware Auditable
             Framework for Large Language Models},
  author  = {Li, D. and Yu, G. and Wang, X. and Liang, B.},
  journal = {Electronics},
  volume  = {15},
  number  = {1},
  pages   = {56},
  year    = {2026}
}

@inproceedings{debenedetti2024agentdojo,
  title     = {AgentDojo: A Dynamic Environment to Evaluate Prompt Injection
               Attacks and Defenses for LLM Agents},
  author    = {Debenedetti, Edoardo and Zhang, Jie and Balunovi{\'c}, Mislav and
               Beurer-Kellner, Luca and Fischer, Marc and Tram{\`e}r, Florian},
  booktitle = {Advances in Neural Information Processing Systems (NeurIPS),
               Datasets and Benchmarks Track},
  year      = {2024}
}

@inproceedings{secalign2025,
  title     = {SecAlign: Defending Against Prompt Injection with Preference
               Optimization},
  author    = {Chen, Sizhe and Zharmagambetov, Arman and Mahloujifar, Saeed and
               Chaudhuri, Kamalika and Wagner, David and Guo, Chuan},
  booktitle = {Proceedings of the ACM SIGSAC Conference on Computer and
               Communications Security (CCS)},
  year      = {2025}
}

@misc{signet2026,
  author = {{Prismer-AI}},
  title = {Signet},
  howpublished = {Open-source software and documentation},
  year = {2026},
  url = {https://github.com/Prismer-AI/signet},
  note = {Accessed September 26, 2026}
}

@techreport{munoz2026permit,
  author = {Munoz, Christian},
  title = {A {SCITT} Profile for Pre-Execution {AI} Action Authorization Records},
  institution = {Internet Engineering Task Force},
  type = {Individual Internet-Draft},
  number = {draft-munoz-scitt-permit-profile-01},
  year = {2026},
  month = jul,
  url = {https://datatracker.ietf.org/doc/html/draft-munoz-scitt-permit-profile-01},
  note = {Work in progress}
}

@techreport{scitt2025architecture,
  author = {Birkholz, Henk and Delignat-Lavaud, Antoine and Fournet, Cedric and
            Deshpande, Yogesh and Lasker, Steve},
  title = {An Architecture for Trustworthy and Transparent Digital Supply Chains},
  institution = {Internet Engineering Task Force},
  type = {Internet-Draft},
  number = {draft-ietf-scitt-architecture-21},
  year = {2025},
  month = sep,
  url = {https://www.ietf.org/archive/id/draft-ietf-scitt-architecture-21.html},
  note = {Work in progress; cited version}
}

@inproceedings{haeberlen2007peerreview,
  author = {Haeberlen, Andreas and Kouznetsov, Petr and Druschel, Peter},
  title = {{PeerReview}: Practical Accountability for Distributed Systems},
  booktitle = {Proceedings of the 21st ACM Symposium on Operating Systems Principles},
  pages = {175--188},
  year = {2007},
  doi = {10.1145/1294261.1294279}
}

@misc{keel2026,
  author = {{Keel}},
  title = {Decision Model and Trusted Action Fields},
  howpublished = {Public software documentation},
  year = {2026},
  url = {https://docs.keelapi.com/governance/decision-model},
  note = {Accessed September 26, 2026; not independently evaluated}
}

@article{saltzer1975protection,
  author = {Saltzer, Jerome H. and Schroeder, Michael D.},
  title = {The Protection of Information in Computer Systems},
  journal = {Proceedings of the IEEE},
  year = {1975},
  url = {https://web.mit.edu/Saltzer/www/publications/protection/}
}

\onecolumn
\appendix
% The detailed related-work comparison is consolidated in Section 8.

\section{Experimental artifacts}
\label{sec:artifact-provenance}

The experiments use two immutable published npm artifacts: CATP 0.7.5
(source \texttt{077c599}) and CATP 0.7.6 (source \texttt{fec7e8e}). Their full
commit identifiers, registry integrity values, lockfiles and installation
inventories are retained with the accompanying material.

Adapter conformance, real-runtime mediation, the original 29-case failure
matrix, enforcement overhead and anchoring measurements use 0.7.5. The
Groth16 study and the 32/32 durability unit results use separately pinned source
companions from that release; the durability source and test bytes are unchanged
between the two releases. The unit results retain their original source identity
and can be reproduced with the companion's locked runner.

The ``Before checks'' and ``After checks'' columns in
Table~\ref{tab:mechanism} refer to 0.7.5 and 0.7.6, respectively. The former
omits five copied-field equality checks; the latter adds them. Both process the
identical signed receipt/export pairs, verified by input hashes. The corrected
artifact meets 13/13 expectations and also passes the repeated 29/29 CLI matrix.
Receipt issuance and verification timings use this corrected artifact, with
100 samples after 10 warmups. The selected run is
\texttt{remediation-0.7.6/run-20260927-01}. These timings are separate from the
original runs and do not support a version-speed comparison. The argv probe
checks both artifacts and finds the same information loss; this limitation is
not fixed by the receipt correction.

Both installations match their registry package bytes. The later inventory gate
checks all 209 installed files/links before execution. Original runs checked the
CLI version and executable hash; they do not retrospectively acquire the stronger
gate. Tables retain source and result identities in their generated source
headers, and raw metadata records dependencies, runtime versions and proof
parameters. Historical failures and the earlier harness run remain available.

\section*{Ethics Statement}
Experiments use isolated local test directories and synthetic tool requests.
The receipt-consistency study does not execute its proposed tool commands, contact
external targets, or use production secrets. Research signing keys are temporary.
The discovered verifier gap and failed cases are retained with their limitations;
no claim of protection against a compromised host or universal prompt injection
is made.

\section*{Open Science Statement}
The companion material contains pinned registry inputs, raw results, table
scripts, a fixed-source durability test companion and the signed counterexamples
with their before/after results. The accompanying source package includes these materials as ancillary files,
with reproduction instructions and an additional adapter-boundary probe. The
CLI source is available at \url{https://github.com/lfzkoala/catp}; the evaluated
npm releases are pinned separately. The results do not describe subsequent
working-tree changes. This manuscript is a preprint, not a peer-reviewed
publication.

\section*{Acknowledgments}
OpenAI Codex assisted with source review, experimental harness code, manuscript
revision and validation. An additional model review was
used for argument and code checks; it is not an independent external peer review.
The author remains responsible for reviewing the evidence and final submission.

\end{document}